\documentstyle[twoside,fleqn,espcrc2]{article}

\title{\bf Cooper pairs as low-energy excitations in the normal state}

\author{O.~Tchernyshyov and Y.~J.~Uemura%
\address
{Physics Department, Columbia University, New York, New York 10027, U.~S.~A.}%
\thanks{This work has been supported by International Joint Research 
Grant from NEDO (Japan) and by the grant DE-FG02-92 ER40699 from DOE 
(USA).}%
}
       
\begin{document}

\begin{abstract}
We discuss the normal state of a fermionic system 
in an idealized {\em pseudogap regime},  
$k_B T_c \leq k_B T \ll |\Delta| \ll \epsilon_F$.
Stable Cooper pairs induce a pseudogap of width $|\Delta|$ in the fermion 
energy spectrum.  Near two dimensions, we find a Bose-like condensation
temperature in this predominantly fermionic system.

\end{abstract}

\maketitle

\section{Motivation}

A number of studies of high-$T_c$ superconductors show that cuprate
superconductors have a short coherence length, comparable 
to the average distance between doped 
holes.  While this observation casts doubt on the applicability of the BCS 
model, the limit of local pairs is not a good description of the cuprates 
either, as the existence of a Fermi surface demonstrates.  A more 
appropriate model of a superconductor with medium-size Cooper pairs 
should feature both fermionic and bosonic degrees of freedom at low energies.

\section{Away from the BCS limit... Whereto?}  

In a Fermi liquid, the propagator of a pair with total momentum ${\bf K}=0$ 
has a cut along the real axis of frequency $\Omega$ (from the continuum of 
free two-fermion states) and 
two poles in the unphysical part of the complex $\Omega$ plane.  As the 
temperature is lowered, the poles approach the real axis in the well-known
\cite{AGD} way: $\Omega \sim \pm i (8/\pi)k_B(T-T_c)$.  
At $T=T_c$, these bound states become stable excitations 
with zero energy and thus immediately undergo Bose condensation.  

To see why stable Cooper pairs in a Fermi liquid {\em necessarily} have zero
energy, consider the dynamical balance between free fermions 
and bound states at ${\bf K}=0$.  Two free fermions with 
opposite momenta and equal energies $\epsilon$
form a bound state with energy $\Omega=2\epsilon$ at a rate
$\Gamma_{+}$ proportional to the fermion occupation 
numbers $n(\epsilon)=(e^{\epsilon/k_B T}+1)^{-1}$ and the density of 
states (DOS): $\Gamma_{+}\propto {\cal D}(\epsilon)[n(\epsilon)]^2$.  
The dissociation rate is 
$\Gamma_{-}\propto {\cal D}(\epsilon)[1-n(\epsilon)]^2$, 
with the same prefactor (by virtue of the time-reversal symmetry).  The 
balance $\Gamma_{+}=\Gamma_{-}$ is achieved when $\epsilon=0$, hence 
$\Omega=0$.  

Note that $\Gamma_{+} - \Gamma_{-} = 0$ is also when the 
DOS vanishes in some interval of energies 
$-|\Delta| < \epsilon < |\Delta|$.  In this case, however, a stable bound 
state can be formed at {\em any} real energy in the interval 
$-2|\Delta|<\Omega<2|\Delta|$
and the condensation is not imminent.  In the BCS model, the origin of 
the energy gap is scattering of fermions from the condensate of pairs,
hence first the pairs condense and then the gap opens up.  However,
when the pair decay is slow (just above $T_c$), scattering 
from these long-lived resonances should already start to modify the fermion 
spectrum.  This observation demonstrates that the question ``which 
comes first, the stable pairs or the gap?'' has no simple answer 
beyond the mean-field treatment.  Quasistable pairs can produce a 
pseudogap, in which they might live long enough; if their energy is 
non zero, they will not form a condensate.  

The BCS model is an idealization that is realized when the Fermi sea 
is robust against the formation of a gap by almost stable Cooper pairs.
Note that the BCS approach is justified only {\em a posteriori} 
(the Ginzburg criterion), when the spectrum of its fluctuations is found.
It would be most advantageous to have a similarly idealized picture
for the opposite limit, when creation of stable pairs precedes 
their condensation.  We also want to stay with a predominantly fermionic 
system, {\em not} a collection of hard-core bosons that arises in 
the limit of fatally strong attraction between fermions.  Whether this
idealization describes any real superconductors ({\em e.g.,} underdoped 
cuprates) is a separate question.  

Some of the features of such an ideal system can be readily anticipated. 
The fact that Cooper pairs have non-zero energies and momenta means 
that each fermion state is no longer coupled to a single hole state 
(by the scattering process ``fermion $\to$ pair + hole''), but rather 
to a continuum of hole states.  For this reason, fermions
have a finite lifetime and the gap is smeared into a pseudogap.  
Nevertheless, if the DOS vanishes in the middle of the 
pseudogap, low-energy Cooper pairs will be almost stable.  The thermal
energy of a boson must be small compared to the 
gap width $|\Delta|$, which itself should not exceed the Fermi energy 
$\epsilon_F$.  We thus define a {\em pseudogap regime}:
\begin{equation}
\label{pseudogap regime}
k_B T_c \leq k_B T \ll |\Delta| \ll \epsilon_F.
\end{equation}

\section{Slowly fluctuating pairing field}
\label{Section Slow}

% Coupling to a continuum and what can be done about it.

The coupling of a fermion to a continuum 
of holes creates a computational problem.  
Its physical consequence is broadening of fermion energy levels
of order $k_B T$.  If, however, there is another mechanism producing
a far stronger broadening, the former effect can be neglected for {\em some}
purposes.  Specifically, we neglect the slight 
difference between the 4-momenta of the fermion and the hole, which simplifies
the problem dramatically.  By doing so, we slightly violate the conservation
of energy and momenta, which may lead to serious errors in computing 
transport properties but has little influence on the fermion propagator.  

% Slowly fluctuating pairing field.  $2+\epsilon$. 

Such broadening can be caused by the pairing field with a fluctuating 
amplitude $|\Delta|$ that shifts 
the energy of a fermion from a bare value $\epsilon$ to 
$\tilde{\epsilon}=\pm(\epsilon^2 + |\Delta|^2)^{1/2}$.  
Amplitude fluctuations of order $|\Delta|$ 
result in the width of order $|\Delta|\gg kT$ for low-energy fermion levels.

% Sadovskii's model. 

This idea can be illustrated by a model in which fermions are coupled to an
external pairing field $\Delta({\bf r},t)$ obeying Gaussian statistics:
\begin{equation}
{\cal H}_{\rm int}({\bf r},t) = 
\Delta({\bf r},t)\psi^\dagger_\uparrow({\bf r},t)
\psi^\dagger_\downarrow({\bf r},t) + {\, \rm H.c.}
\end{equation}
When temporal and spatial fluctuations of $\Delta({\bf r},t)$ 
are neglected, $\langle\Delta({\bf r},t)\Delta^*(0,0)\rangle = |\Delta|^2$, 
the problem admits an exact solution \cite{Sad74}.  
Low-energy fermion excitations have the linewidth of order $|\Delta|$.
The DOS exhibits a pseudogap of width $|\Delta|$ 
and vanishes quadratically at low energies.

\section{Cooper pairs in 2+$\varepsilon$ dimensions}
%{Self-consistent T-matrix approach}

% T-matrix approach. 

The pseudogap regime (\ref{pseudogap regime}) may be realized in 
quasi two-dimensional superconductors with moderately weak attraction
between fermions, where $T_c$ is reduced substantially 
below the BCS value $T_c^{\rm BCS}\approx |\Delta|$.  
(We do not consider 
here the possibility of a Kosterlitz-Thouless transition.)  
We have studied the pseudogap regime in $2+\varepsilon$ dimensions using
the self-consistent T-matrix approach \cite{Baym}, which readily permits
introduction of a propagator for Cooper pairs \cite{Oleg}.  The
resulting set of diagrams for the fermion propagator is similar to the 
Gaussian model of the previous section, but without crossing boson lines.  
A smaller number of diagrams leads to a weaker suppression of the fermion
states, ${\cal D}(\epsilon)\propto \epsilon$ at low energies; otherwise,
the DOS has a similar structure.  

% Cooper pair propagator. 

Inside the pseudogap, long-lived Cooper pairs have the dispersion of a
Bogoliubov mode with a mass, which prevents their condensation: 
\begin{equation}
\Omega^2 = \Omega_0^2 + s^2{\bf K}^2,
\end{equation}
where $s^2=\epsilon_F/m$.  The mass term $\Omega_0$ is determined 
self-consistently to match the pseudogap width with 
the average fluctuation of the pairing field.  
In terms of the fermion density $n$ and mass $m$, the resulting condensation 
temperature is 
\begin{equation}
k_B T_c \sim (\varepsilon\pi/6)\hbar^2 n/m = (2/3)k_B T_B,
\end{equation}
where $T_B$ is the condensation temperature of an ideal Bose gas with 
density $n/2$ and mass $2m$.

% Self-consistency and $T_c$.

\end{document}